\documentclass{article}
\usepackage{amsmath}
\usepackage{graphicx}
\usepackage{subcaption}
\usepackage{psfrag}
\usepackage{setspace}
\def\fakebold#1{\relax\ifvmode\leavevmode\fi%
\ifmmode%
\setbox0=\hbox{$#1$}%
\else%
\setbox0=\hbox{#1}%
\fi%
\kern-.02em\copy0 \kern-\wd0%
\kern .04em\copy0 \kern-\wd0%
\kern-.0125em\raise.02em\box0%
}%

\newcommand{\fatbeta}{\mbox{\fakebold{$\beta$}}}

\begin{document}
\title {The Universal model and prior: multinomial GLMs}
\author{Murray Aitkin, School of Mathematics and Statistics\\
 University of Melbourne}
\maketitle
\begin{abstract}
This paper generalises the exponential family GLM to allow arbitrary 
distributions for the response variable.
This is achieved by combining the model-assisted regression approach from survey 
sampling with the GLM scoring algorithm, weighted by random draws from the 
posterior Dirichlet distribution of the support point probabilities of the 
multinomial distribution.
The generalisation provides fully Bayesian analyses from the posterior sampling, 
without MCMC.
Several examples are given, of published GLM data sets.
The approach can be extended widely: an example of a GLMM extension is given.
\end{abstract}

\section{Introduction}
Discussions of the foundations of survey sampling were extensive in the 1960s 
and 70s, from many points of view.
The following quotes are particularly relevant to the contribution of this paper. 
\begin{itemize}
\item 
The basic question to ask is why should finite population inference be different 
from inferences made in the rest of statistics? 
I have yet to find a satisfactory answer. 
My view is that survey statisticians should accept their responsibility for 
providing stochastic models for finite populations in the same way as 
statisticians in the experimental sciences.
These models can then be treated within the framework of conventional theories 
of inference.
The problems with the Neyman approach then disappear to be replaced by disputes 
between frequentists, Bayesians, empirical Bayesians, fiducialists and so on.
But at least these disputes are common to all braches of statistics and sample 
surveys are no longer seen as an outlier.
Smith (1976)

\item All actual sample spaces are discrete, and all observable random variables 
have discrete distributions.
The continuous distribution is a mathematical construction, suitable for 
mathematical treatment, but not practically observable.
Pitman (1979, p. 1).

\item  The basic feature of our theory is a special parametrization of finite
populations of $N$ units.

... we assume, with essentially no loss of generality, that this characteristic 
[$y$] is measured on a known scale with a finite set of scale points 
\mbox{$y_t (t=1,...,T)$}.
... Any finite population can then be completely described by the set of $T$ 
non-negative integer parameters $N_t$, being the number of units in the 
population having the characteristic $y_t$ and satisfying the condition 
$\sum_{t=1}^T N_t = N$.

... If suitable prior information is available, Bayesian concepts can be 
adjoined to our theory using the complete likelihood. 
...
It will be seen that with our theory every sample design ... requires the 
derivation of its appropriate likelihood for the observables $n_t$.
In this paper the only sampling procedures considered are simple random sampling 
with equal probabilities with or without replacement.
Extensions to multi-stage designs, unequal probability sampling etc., will be 
considered in subsequent papers. 
(Hartley and Rao 1968, pp. 548-9)

\end{itemize}

The model-based and design-based schools of inference have been slowly 
converging.
In the rigid design-based ``model-assisted" approach, set out in detail in 
S\"{a}rndal, Swensson and Wretman (1992), models had a very limited role, 
relating only the population mean and variance parameters to ``auxiliary" 
variables.
Analysis was through the sample selection probabilities and survey weights from 
the inverse selection probabilities.
Least squares was invoked for optimality in model fitting.
 
This constraint on the use of models has been
relaxed in some modern survey 
sampling treatments.
Chambers, Steel, Wang and Welsh (2012) used explicit probability models for 
response variables and their maximum likelihood analysis. 
The survey weights were barely mentioned: analysis was through the ``missing 
information principle", regarding the unsampled part of the population as 
missing data, which was effectively imputed from the model and data of the 
observed sample. 
With non-informative ignorable survey designs, the sample selection indicators 
were ancillary and served no inference function.

The early developments in Fisherian model-based analysis, relying heavily on the 
Central Limit Theorem for asymptotic optimality, were developed much further by 
the GLM (Nelder and Wedderburn 1972) and EM algorithm (Dempster, Laird and Rubin 
1977) inventions.
Their reliance on specific probability models required the further development 
of model evaluation and assessment methods for the exponential family.
Quasi-likelihoods (Wedderburn 1974) were an attempt to extend GLM properties to 
unspecified data distributions.
Bayesian MCMC extensions of EM, beginning with the Data Augmentation algorithm 
(Tanner and Wong 1987) corrected the optimism of confidence intervals through 
credible intervals, which accounted for skewness in the likelihoods for models 
outside the Gaussian (Aitkin 2018 gives simple examples).
However they were equally dependent on the validity of the probability model 
assumption.  
This placed both Fisherian frequentists and Bayesians in the same difficulty as 
the survey samplers: the Fisherian sufficiency and optimality of the likelihood 
for inference depended on the validity of the probability model assumption, but 
this could never be {\em proved correct} -- it could at most be {\em consistent} 
with the data. 

The possible use of the multinomial distribution and its conjugate Dirichlet 
prior as a {\em general} distribution and prior for data analysis, was begun by 
Hartley and Rao (1968), Ericson (1969) and Hoadley (1969).
These papers were necessarily theoretical, since the computational facilities 
needed were not then developed for either the profile likelihood analysis for 
maximum likelihood, or the posterior sampling for Bayesian analysis.

Lindsey (1971, 1974a, 1974b, 1997), Lindsey and Mersch (1992) took the 
multinomial in a different direction, as a basis for {\em modelling} the 
underlying density or mass function, by expressing the multinomial as a set of 
constrained Poisson counts, and using the log-linear Poisson model to fit 
functions of the response variable as model terms.  

Rubin (1981) extended the Bayesian analysis to the {\em non-informative} Haldane 
(1948) Dirichlet prior, and gave it the name {\em Bayesian bootstrap}.
Maximum likelihood analysis was stimulated by the work of Owen (1988) on 
{\em empirical likelihood}.
Maximising the likelihood over the multinomial parameters, constrained by the 
fixed population mean, generated the {\em profile empirical likelihood}.
Owen's book (2001) emphasised the frequentist applications of profile empirical 
likelihoods for the population mean, while recognising the Bayesian extension 
with the conjugate Dirichlet prior.
Guti\'{e}rrez-Pena and Walker (2005) and Walker and Guti\'{e}rrez-Pena (2007)
argued for the multinomial/Dirichlet as the fundamental inference model and 
prior.
The difficulty was extending it to the common useful models, like GLMs and their 
extensions, and to survey designs more complex than the simple random sample.

Aitkin (2008) extended the Bayesian bootstrap to both multiple regression models 
and stratified and clustered survey designs.
A further extension was given there to the regression model parameters in a 
two-level survey design.
It was unclear how to deal with GLMs, since the usual parameters of interest are 
in the linear predictor, whose ML estimates are not linear functions of the 
observations.

J. Rao and his colleagues (for example Wu and Rao 2010 and Yi, Rao and Li 2016) 
combined the empirical profile likelihood with a flat prior on the mean to 
produce a {\em composite} or {\em pseudo} likelihood, analysed in the 
conventional survey sampling framework.

Huang (2014, Zhang and Huang 2018) extended the profile empirical likelihood to 
regression models.
The computations were complex, and implemented only in MATLAB. 
The extension did not include more general sample designs.

The present paper provides a way to express the population parameters of 
interest through the converged scoring algorithm of the GLM analysis, and to 
combine this with the extended Bayesian bootstrap.
This provides a full Bayesian analysis of the GLM, without the usual probability 
model assumption, and without the need for MCMC analysis: simple simulations 
from the Dirichlet posterior distribution are all that is needed.

The paper describes the procedure in Section 2, and subsequent sections give 
examples of increasingly complex GLMs and their analyses.
Section 6 gives discussion of the approach.    

\section{Summary of the procedure}

The procedure involves:
\begin{itemize}
\item a non-informative ignorable survey design;
\item a structural model specification of the population parameters of interest
(the {\em fixed part} of the GLM);
\item a multinomial distribution for the response variable with population 
proportion parameters on the distinct joint support points of the population 
response and covariates;
\item the non-informative Haldane Dirichlet prior on the multinomial parameters;
\item a maximum likelihood (ML) algorithm based on a tentative specification of 
the {\em random part} of the GLM which allows for weighting of the observations.
\end{itemize}
The structural model is fitted by the ML algorithm with a sequence of random 
weights drawn from the posterior Dirichlet distribution of the multinomial 
parameters.
The random weights induce random values of the parameter MLEs, which define the 
posterior distribution of the model parameters.

\section{Example 1: a population mean}

We use an example from Aitkin (2010 Chs 1 and 4) of a simple random sample of 
size 40 from a population of 648, given in Table 1. 
The question of interest is the population mean family income. 

\begin{table}[h]
\caption{Family income data, in units of 1000 dollars}
\centering
\begin{tabular}{|r r r r r r r r r r|}
\hline
26 &35 &38& 39& 42& 46&  47 &47 &47 &52\\ 
53 &55 &55 &56& 58 &60&  60 &60 &60 &60 \\
65 &65 &67 &67& 69& 70&  71 &72 &75 & 77 \\
80 &81 &85 &93& 96& 104& 104& 107& 119& 120\\
\hline
\end{tabular}
\end{table}
\label{income_data_ch4}

\subsection{Multinomial analysis}

For the multinomial analysis, the income population consists of $N$ values 
$Y_I^*$.
We tabulate them conceptually into the $D$ {\em distinct} values $Y_J$ with 
frequency $N_J$.
The probability that a randomly drawn sample value gives the value $Y_J$ is 
$p_J = N_J/N$.
Our interest is not in the $p_J$ but in the population mean
$\mu = \sum_{J=1}^D p_J Y_J.$

The likelihood of the sample is (omitting the known constant)
$$L({\bf p}) = \prod_{J=1}^D p_J^{n_J}.$$
We tabulate the sample values correspondingly, obtaining $d$ distinct values 
$y_j$ with frequencies $n_j$ in Table 2.
\begin{table}[h]
\caption{Income data tabulation}
\centering
\begin{tabular}{|l| r r r r r r r r r r |}

\hline
  $j$ & 1&  2&  3&  4&  5&  6&  7&  8&  9& 10\\
$y_j$ &26& 35& 38& 39& 42& 46& 47& 52& 53& 55\\
$n_j$ & 1&  1&  1&  1&  1&  1&  3&  1&  1&  2\\
\hline
  $j$ & 11& 12& 13& 14& 15& 16 & 17& 18& 19& 20\\
$y_j$ & 56& 58& 60& 65& 67& 69 & 70& 71& 72& 75\\
$n_j$ &  1&  1&  5&  2&  2&  1 &  1&  1&  1&  1\\
\hline
 $j$ & 21& 22& 23& 24& 25& 26&  27&  28&  29&  30\\
$y_j$& 77& 80& 81& 85& 93& 96& 104& 107& 119& 120\\
$n_j$&  1&  1&  1&  1&  1&  1&   2&   1&   1&   1\\
\hline
\end{tabular}
\end{table}
\label{incomedata}

\noindent We use the Haldane Dirichlet D({\bf 0}) prior with $a_J = 0$ for all 
$J$, giving the Dirichlet posterior D({\bf n}), now defined on the $d$ distinct 
values in the observed support: 
$$\pi(p_1,\dots,p_d\,|\, {\bf y}) = \frac{\Gamma(n)}{\prod_{j=1}^d \Gamma(n_j)}
\prod_{j=1}^d p_j^{n_j - 1}.$$
\noindent Population values unobserved in the sample are given zero posterior 
probability, and can be omitted from consideration.

\begin{figure}[h!]
\centering
\includegraphics[width=3in]{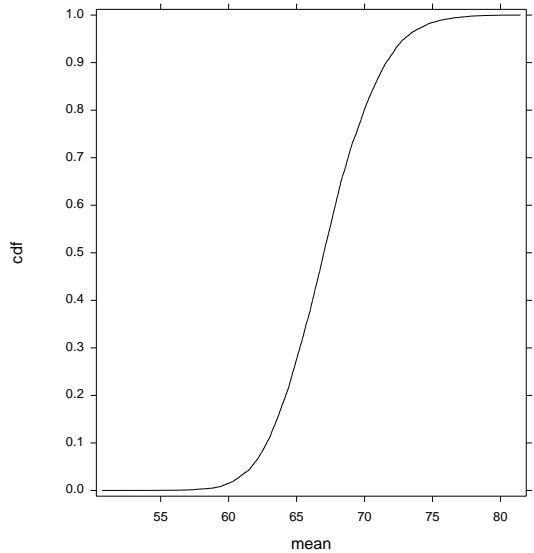}
\caption[Ch2_29]{Posterior cdf, income mean}
\label{incomesimcdf}
\end{figure}

\begin{figure}[ht!]
\centering
\includegraphics[width=3in]{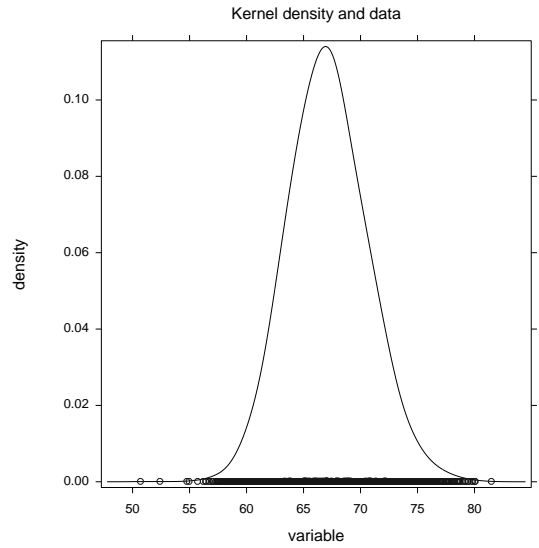}
\caption[Ch2_29]{Posterior density, income mean}
\label{incomesimkern}
\end{figure}

\noindent The posterior distribution from 10,000 draws is shown as a cdf in 
Figure~\ref{incomesimcdf} and as a kernel density, together with the data 
(unjittered) in Figure~\ref{incomesimkern}.
It is slightly right-skewed.
The 95\% central credible interval is [60.6, 74.2].

\begin{figure}[h!]
\centering
\includegraphics[width=3in]{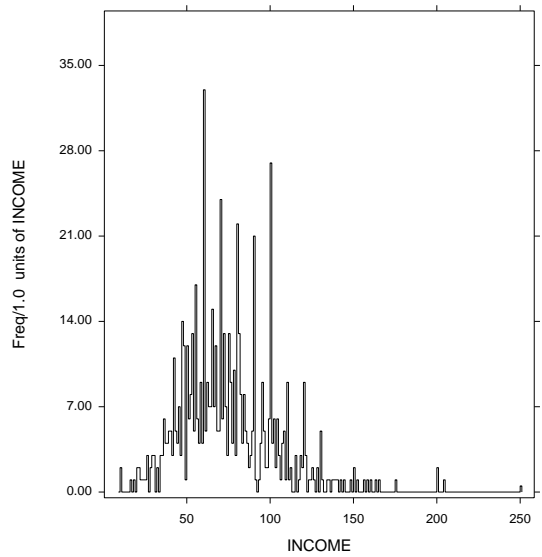}
\caption[Ch2_29]{Income histogram}
\label{popincomehist}
\end{figure}

The sample mean is \mbox{$\bar y = 67.1$} and the (unbiased) variance is 
$s^2 = 500.87.$
The survey-sampling large-sample 95\% confidence interval for the mean is 
\mbox{$\bar y \pm 1.96 s/\sqrt{n}$}, which is $[60.1,74.0]$; this is nearly 
identical to the $t$-interval $[59.9, 74.3]$ assuming a normal distribution for 
income.
The design-based interval using the finite population correction of 
$(1 - 40/648) = 0.938$ gives the slightly shorter interval $[60.6,73.6]$.
These intervals are all in close agreement, despite the unusual shape of the 
population, in Figure \ref{popincomehist}.
It is far from smooth or well-represented by a gamma or lognormal model.

\psfrag{Freq/1.0 units of INCOME}{frequency}

\section{GLM formulation}

An important point for the GLM is that the above analysis can be extended 
directly to multiple regression models, and can be expressed as a 
posterior-weighted form of the ML analysis, assuming the structural model for 
the mean.
For a sample $(y^*_i,{\bf x}^*_i)$ of size $n$, we model the joint 
distribution as a multinomial with probabilities $p_J$ on the {\em distinct} 
population values $(Y_J,{\bf X}_J)$.
With the non-informative Dirichlet prior, we again obtain the posterior on the 
{\em observed} distinct sample support values $(y_j, {\bf x}_j)$.
We {\em define} the population parameter of interest as the {\em population} 
value 
$${\bf B}=\mathcal X'\mathcal\mathcal X]^{-1} \mathcal X'{\bf Y}$$,
where $\mathcal X$ is the population matrix of covariate values and ${\bf Y}$ 
the population dependent variate.
We make $M$ draws $p_j^{[m]}$ for each distinct observation and take them as 
{\em prior weights} in the GLM sense, leading to $M$ ML estimates of the 
regression coefficient vector: 
$$\boldsymbol{\beta}^{[m]} = [X'W^{[m]}X]^{-1} X'W^{[m]}{\bf y},$$
where W is the diagonal matrix of the Dirichlet posterior draws $p_j$.
These estimates define the posterior distribution of $\boldsymbol\beta$.

We do not need to investigate the residual distribution to check its 
specification: the population parameters of interest have been defined by the 
user, and the posterior weighting provides protection against mis-specification 
of a Gaussian distribution.
But this still provides an efficient analysis in the Fisherian sense: we have 
used the likelihood, and the ``nonparametric" multinomial and prior provide the 
{\em minimal} information necessary for a posterior distribution statement.

This approach can be extended to general GLMs with an arbitrary response 
distribution: we need only to specify the population model parameters of 
interest through the GLM representation. 

\section{Example 2: vaso constriction}

Finney's data (1947) on vaso constriction in the skin of the digits of the hand 
are used widely in statistical packages and books (for example Aitkin, Francis, 
Hinde and Darnell 2009) as an example of logistic regression.
The response variable is 39 measures of the presence (1) or absence (0) of the 
vaso constriction response in subjects; the covariates are the values of 
volume and rate of air inspired.
We fit by ML a logistic regression with variables log volume (LV) and log rate
(LR).
The estimates and SEs (standard errors) are :
$$\widehat{\beta_0} = -2.88 (1.32), \, \widehat{\beta_{LV}}= 5.18 (1.86),\,
\widehat{\beta_{LR}} = 4.56 (1.84).$$ 

For the multinomial model, we need to define the population regression 
parameters of interest.
S\"{a}rndal et al (1992) do not deal with GLMs.
We give a general definition of the GLM population model parameters, as 
population analogues of the IWLS scoring algorithm; in this algorithm we write 
at the $r$-th iteration:
$$\fatbeta_{r+1} =  [X'W_rX]^{-1} X'W_r{\bf z_r},$$ 
where $W$ is the matrix of iterative weights and ${\bf z}$ is the adjusted 
dependent variate; both are functions of the model parameters.  
At convergence of the algorithm, we have 
$$\widehat{\fatbeta} =  [X'\widehat{W}X]^{-1} X'\widehat{W}{\bf \widehat{z}},$$
where $\widehat{W}$ is the matrix of converged iterative weights and 
$\widehat{\bf z}$ is the converged adjusted dependent variate
We define the population regression parameters $B$ by
$$ B = [\mathcal X' \mathcal W \mathcal X]^{-1} \mathcal X \mathcal W {\bf Z},$$
where $\mathcal X$ is the population matrix of covariate values, $\mathcal W$ 
the population matrix of weights, and ${\bf Z}$ the population adjusted 
dependent variate.
(Of course these are not observable.)

We adapt this definition to generalise the ML algorithm above.
We use the IWLS algorithm with additional weighting of the weight matrix by the
random draws of the Dirichlet posterior probabilities $p_j^{[m]}$  on the 
observed support.
So the IWLS algorithm for the GLM ML estimation is randomly iteratively weighted 
as in the previous example: we have at convergence and the $m$-th draw

$$ \fatbeta^{[m]} = [X'W^{[m]}X]^{-1} X'W^{[m]}{\bf z}.$$

\begin{figure}[hb!]
\centering
\includegraphics[width=3in]{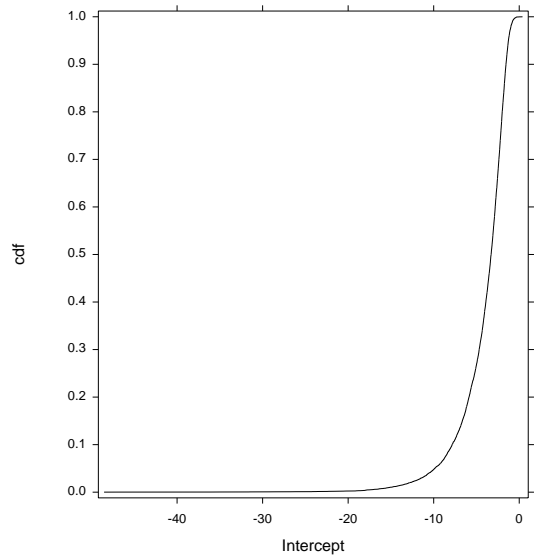}
\caption{Vaso-constriction: intercept posterior}
\label{vasoalpha}
\end{figure}

\begin{figure}[ht!]
\centering
\includegraphics[width=3in]{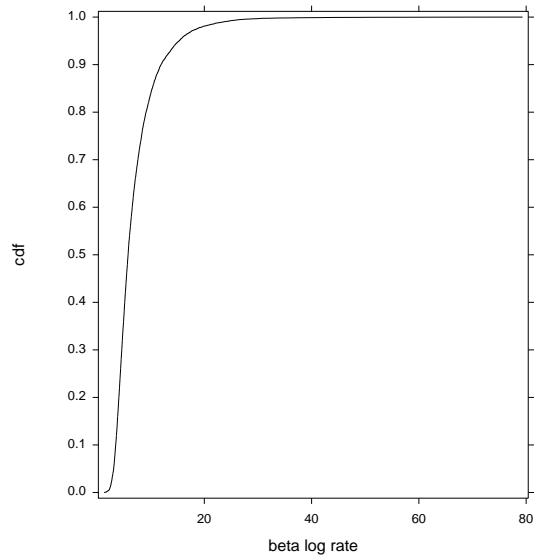}
\caption{Vaso-constriction: log rate slope posterior}
\label{vasobetar}
\end{figure}

\begin{figure}[h!]
\centering
\includegraphics[width=3in]{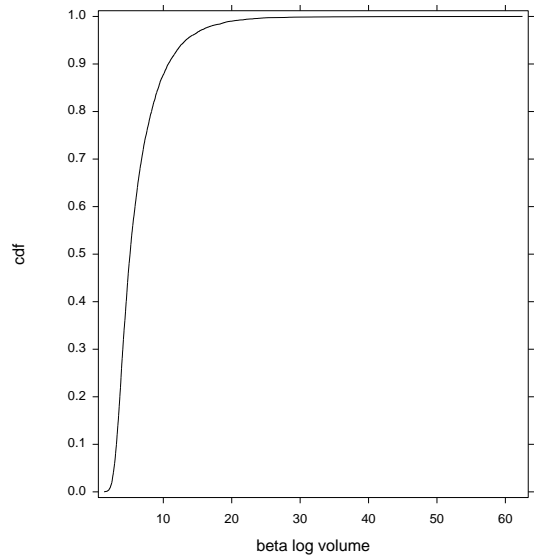}
\caption{Vaso-constriction: log volume slope posterior}
\label{vasobetav}
\end{figure}

\noindent The posterior distributions of the parameters from $M$=1,000 draws are 
shown in Figures~\ref{vasoalpha}, ~\ref{vasobetar} and ~\ref{vasobetav}.
They are {\em very} heavily skewed.

\begin{figure}[ht!]
\centering
\includegraphics[width=3in]{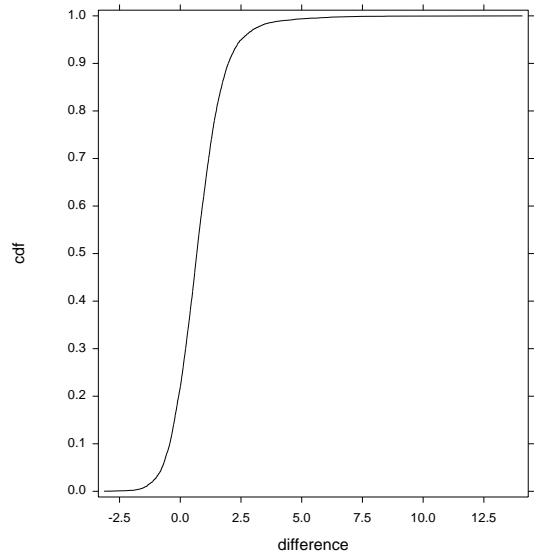}
\caption{Vaso-constriction: log volume - log rate posterior}
\label{vaso_difference_posterior}
\end{figure}

The similarity of the two regression coefficients has suggested that they could 
be equated.
To assess the plausibility of this, we show in 
Figure~\ref{vaso_difference_posterior} the posterior cdf of $\beta_{LV} - 
\beta_{LR}$.
The central 95\% credible interval includes zero.
We proceed with the model of a common regression coefficient $\beta$ on the 
composite variable $LT = LR+LV$, with intercept $\alpha$.
The ML estimates and (SE)s are
\mbox{$\widehat\alpha = -3.05 (1.27) $} and \mbox{$\widehat\beta = 4.93 (1.72).$} 

\begin{figure}[ht!]
\psfrag{lt}{LT}
\centering
\includegraphics[width=3in]{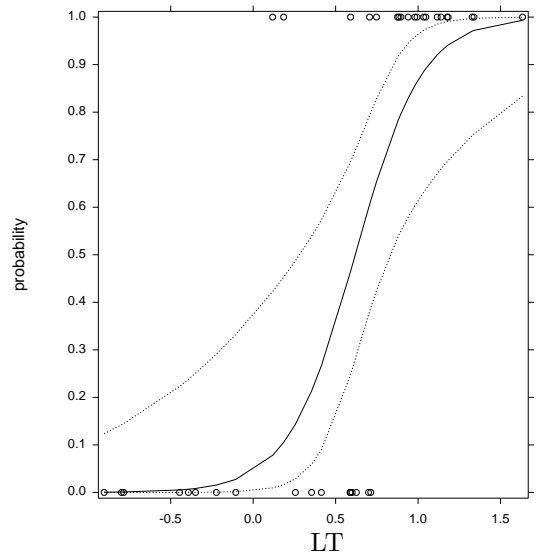}
\caption{Vaso-constriction: data, ML fitted LT model (solid curve) and 95\% 
confidence region (interior of dashed curves)}
\label{vaso_ML_fitted_bounds}
\end{figure}

We compare the fitted models with the composite variable $LV+LR$, and their 
precisions, by ML and posterior weighted ML.
Figure~\ref{vaso_ML_fitted_bounds} shows the ML fitted model (solid curve) and 
the 95\% confidence region, computed on the logit scale and transformed, based 
on the information matrix (dashed curves).
The confidence region is very wide: the sample of 39 is too small for any 
precision.

\begin{figure}[ht!]
\centering
\psfrag{alpha}{$\alpha$}
\includegraphics[width=3in]{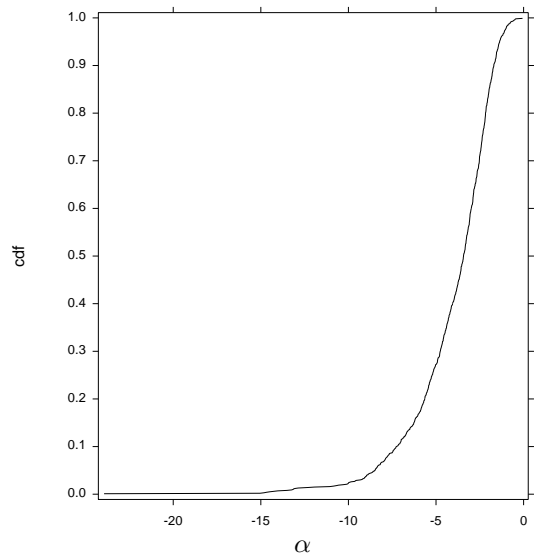}
\caption{Vaso-constriction: posterior distribution of $\alpha$ for common 
$\beta$}
\label{vaso_alpha_common}
\end{figure}

\begin{figure}[h!]
\centering
\psfrag{ B }{$\beta$}
\includegraphics[width=3in]{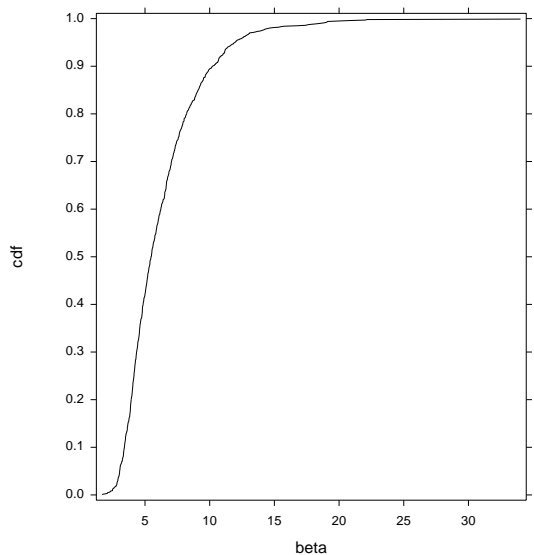}
\caption{Vaso-constriction: posterior distribution of common $\beta$}
\label{vaso_beta_common}
\end{figure}

Figures~\ref{vaso_alpha_common} and ~\ref{vaso_beta_common} show the posterior 
distributions of $\alpha$ and $\beta$.

\eject
\noindent The posterior median and 95\% central credible intervals are: for 
$\alpha$, $-3.43$ and $[-9.94, -1.07]$, and for $\beta$, 5.46 and [2.84, 13.92].
The posterior medians are larger in magnitude than the MLEs: the Bayesian median 
curve has moved slightly to the right, and has a slightly steeper slope than the 
ML curve.
The 95\% confidence intervals: $[-5.59, -0.51]$ for $\alpha$ and [1.49, 8.37] 
for $\beta$ are much shorter and are mislocated: the covariance matrix of the ML 
estimates cannot represent or allow for the severe skewness in the parameter 
posteriors.

\begin{figure}[ht!]
\centering
\includegraphics[width=3in]{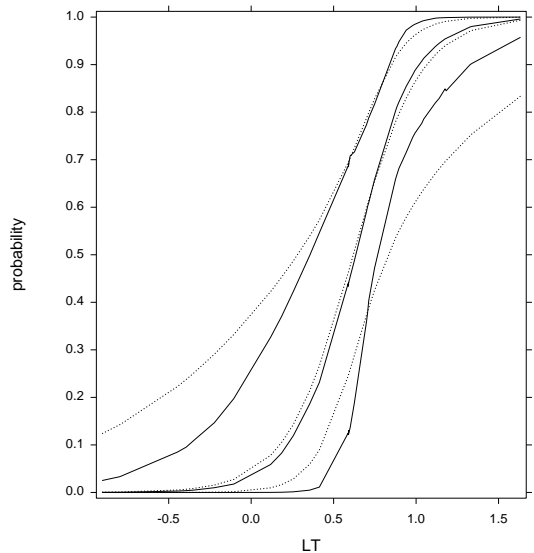}
\caption{Vaso-constriction: fitted ML (dashed curves) and Bayes (solid curves) 
models and 95\% bounds}
\label{vaso_Bayesbounds2}
\end{figure}

Figure~\ref{vaso_Bayesbounds2} shows both the fitted ML logistic regression with 
the bounds of the 95\% confidence region (dashed curves) and the posterior 
median of the fitted model, with the bounds of the 95\% credible region (solid 
curves) from 10,000 draws.
The differences between the two sets of bounds are greater than those between 
the medians, and these differences increase away from the 50\% probability, 
especially for low values of LT.

\noindent 
\section{Example 3: absence from school}

A demanding data set on children's absence from school was discussed in 
Aitkin (1978) and subsequently in Aitkin, Anderson, Francis and Hinde (1989) and 
in other books and articles (for example Venables and Ripley 2002).
The data, from 146 children, are counts of days absent from school and form a 
4-way unbalanced cross-classification by culture C, sex S, age group A and 
normal/slow learner L.
There were eight zero values of absence.
The full data were given in the discussion of Aitkin (1978 p. 223), and can be 
found in R.

The analysis in Aitkin (1978) used Gaussian-theory-based ANOVA for the 
unbalanced cross-classification. 
The second analysis in Aitkin (1978) using a lognormal distribution of (days+1) 
established the importance of the three-way CSL interaction, and the irrelevance 
of the four-way and other three-way interactions.
  
Aitkin et al (1989) considered a Poisson analysis for the counts; the log-linear 
Poisson model scale would give similar parameters to the lognormal analysis.
They pointed out however the severe overdispersion of values within many cells 
of the classification, invalidating the Poisson standard errors.
They concluded that neither a negative binomial nor a quasi-likelihood analysis 
was appropriate; as Nelder said in his 1978 discussion 

``These data are intrinsically awkward."

We define the population parameters of interest as those from the population 
version of the converged Poisson log-linear model.
The previous analyses used the full set of degrees of freedom in each model term.
To allow for the possibility of one or more cells departing in mean absence from 
a simple model for the others, we convert the classification variables to 
dummies, and regard all the cross-product terms constructed from these as 
potential candidates for omission from the ``full" model.
This approach allows for the omission of ``main effect" dummies while retaining 
``interactions" between the dummies. 

We fit a sequence of  models, beginning with the ``full model" of CSL and its 
marginal main effects and interactions, and all main effects and two-way 
interactions of the other three factor dummy variables.
We use model reduction by backward elimination, dropping from each model the 
dummy variable with the smallest ratio of posterior mean to posterior standard 
deviation -- the equivalent of a $t$-statistic in ML variable elimination.
(The parameter posterior distributions in all the models considered were 
symmetric and close to Gaussian.)
Model reduction is speeded-up by recycling the fitted values from each model fit 
as the starting values for the next model fit.

We generate 1000 random draws from the posterior Dirichlet distribution on the 
observed support, and weight the ML parameter estimation by these draws, 
generating 1000 random draws of the posteriors of all the parameters in each 
model.

Variable elimination proceeds as it would for the Poisson model assuming the 
standard errors were correct, but continues further, as the weighted parameter 
ML estimates are similar, but the standard errors from the Poisson model 
understate the posterior standard deviations by factors of 2--3, different for 
each variable.

Reduction terminates with 11 model variables, by a criterion of ratio of mean to 
standard deviation greater than 2 (the smallest remaining was 2.65, the next 
3.47).
The final model posterior means (pmeans) and standard deviations (psd) are given 
in Table 3.

\begin{table}[h]
\centering
\begin{tabular}{|ccccccc|}
\hline
variable& 1&C  &S   &CS   &CL   & CSL\\
\hline
pmean&2.17&1.01&0.78&-1.20&-1.13&1.35\\
psd  &0.17&0.34&0.22& 0.30& 0.32&0.35\\
\hline
\hline
variable& A3&  A4&  CA2&CA3  &SA3  &  SA4\\
\hline
pmean&1.29&1.17&-0.99&-1.25&-0.85&-1.53\\
psd  &0.21&0.21& 0.29& 0.33& 0.32& 0.32\\
\hline
\end{tabular}
\caption{Final model parameters}
\end{table}
\label{absence_final_model}

\noindent This 11-parameter model is both more and less complex than that given 
in the discussion of Aitkin (1978).
It is more complex, in having important components of the CA interaction as well 
as of the CSL and SA interactions, but it is less complex in omitting 
unimportant components of the latter interactions.

The following tables give the posterior median values (to 1 dp, upper value) and 
the observed means (rounded) and sample sizes (lower values in parentheses) of 
days absent by school level for each cell in the cross-classification. 

{\bf Primary, slow learners}
\begin{table}[h]
\centering
\begin{tabular}{|cccc|}
\hline
Aboriginal&      & White &\\
\hline
Girls     & Boys & Girls & Boys\\
\hline
8.8       & 19.1   & 24.0 & 14.9\\
(9,3)   & (3,1)&(30,3)& (25,1)\\
\hline
\end{tabular}
\end{table}

{\bf Primary, average learners}
\begin{table}[h]
\centering
\begin{tabular}{|cccc|}
\hline
Aboriginal&      & White &\\
\hline
Girls     & Boys & Girls & Boys\\
\hline
8.8       & 19.1   & 7.8 & 19.7\\
(13,5)   & (21,4)&(5,6)& (18,4)\\
\hline
\end{tabular}
\end{table}

\noindent For average learners, white and Aboriginal children have the same 
pattern of absence -- boys are absent more than twice as often as girls.
This is also true for Aboriginal slow learning girls, but white slow learning 
girls are absent nearly twice as often as boys.

{\bf Secondary 1, slow learners}
\begin{table}[h]
\centering
\begin{tabular}{|cccc|}
\hline
Aboriginal&      & White &\\
\hline
Girls     & Boys & Girls & Boys\\
\hline
8.8       & 19.1   & 8.9 & 5.5\\
(9,3)   & (23,10)&(6,7)& (6,11)\\
\hline
\end{tabular}
\end{table}

{\bf Secondary 1, average learners}
\begin{table}[h!]
\centering
\begin{tabular}{|cccc|}
\hline
Aboriginal&      & White &\\
\hline
Girls     & Boys & Girls & Boys\\
\hline
8.8       & 19.1   & 2.9 & 7.3\\
(10,2)   & (11,5)&(3,2)& (11,6)\\
\hline
\end{tabular}
\end{table}

\noindent The pattern of absence in first year for Aboriginal chldren is the 
same as for those in the last primary year.
For white children, slow learning girls are absent as often as Aboriginal girls, 
and three times as often as average learning girls, while absence for white boys 
is nearly the same for slow and average learners.
\eject 
{\bf Secondary 2, slow learners}
\begin{table}[h!]
\centering
\begin{tabular}{|cccc|}
\hline
Aboriginal&      & White &\\
\hline
Girls     & Boys & Girls & Boys\\
\hline
31.8      & 13.6 & 25.0  & 10.7\\
(37,4)   & (36,8)&(29,3)& (6,9)\\
\hline
\end{tabular}
\end{table}

{\bf Secondary 2, average learners}
\begin{table}[h!]
\centering
\begin{tabular}{|cccc|}
\hline
Aboriginal&      & White &\\
\hline
Girls     & Boys & Girls & Boys\\
\hline
31.8      & 13.6   & 25.0 & 10.7\\
(27,7)   & (2,1)&(9,7)& (1,1)\\
\hline
\end{tabular}
\end{table}

\noindent Slow and average learners have the same pattern of absence. 
Girls are absent 2.5 times as often as boys, and Aboriginal children are absent  
25\% more often than white children.

{\bf Secondary 3, average learners}
\begin{table}[h]
\centering
\begin{tabular}{|cccc|}
\hline
Aboriginal&      & White &\\
\hline
Girls     & Boys & Girls & Boys\\
\hline
28.2      & 6.1   & 25.0 & 5.4\\
(27,7)   & (15,9)&(27,7)& (13,10)\\
\hline
\end{tabular}
\end{table}

\noindent Absence is nearly the same for Aboriginal and white children, but  
girls are absent nearly five times as often as boys.

The overall conclusions are similar to those in Aitkin (1978), but there are 
several differences, due to both the different final models and the extended 
variable elimination from the much larger variabilities of the parameters.

\section{Discussion}

Bayesian applications of empirical likelihood are few. 
Most of the applications in Owen (2000) are frequentist, and recent work by 
Huang (2017) and Zhang and Huang (2018) follows the same path. 
Applications to GLMs are complicated by the optimisation problem, and few 
general-purpose algorithms are available.

A few Bayesians, notably Guti\'{e}rrez-Pena and Walker, have argued for the 
multinomial/Dirichlet combination as a {\em general} model and prior for data 
analysis.
Rao and his survey colleagues (Rao and Wu 2010a and 2010b, Wu and Rao 2010, and
Datta, Rao and Torabi 2010) have combined the empirical profile likelihood with 
a flat prior on the mean to develop a composite ``Bayesian pseudo-empirical 
likelihood" approach. 

The Bayesian bootstrap posterior weighting approach makes a valuable 
contribution to all three schools of statistical inference.
Each school is effective within its box, but we are now able to look outside the 
boxes.
\begin{itemize}
\item The Bayesian bootstrap posterior weighting approach resolves the very 
long-term argument over the role of models in the design-based approach.
The old argument that official statistics reporting is too important to rely on 
possibly (or {\em inevitably}) incorrect probability models can now be inverted.
The multinomial model provides an {\em always true} model with {\em efficient}
inference through the likelihood and posterior distribution of the user's 
specified model parameters.
The ancillary sample selection indicators are no longer needed for inference 
with non-informative survey designs.
The new argument is that official statistics reporting is too important to rely 
on possibly (or {\em inevitably}) incorrect precision statements from standard 
errors, robust or not. 

\item The skewness of likelihoods outside the Gaussian, ignored in the classical 
asymptotic ML theory, is fully recognised and allowed for: the understatement of 
variability and location in symmetric confidence intervals is corrected. 
At the same time the computational value of maximum likelihood in inference is 
increased: ML estimates, for example from the EM algorithm, are sufficient.
We do not need their standard errors. 

\item The ``nonparametric" Bayesian bootstrap is fully generalised to handle any 
non-informative survey design and any specific structural model.
Model-dependent MCMC methods are not needed, and would not in any case account 
for departures from the probability model assumption.
\end{itemize} 

Several Bayesian commentators have argued -- 
\begin{quote}
The discrete nature of the multinomial/Dirichlet is limiting, outdated and 
unnatural.  
Why not use the Dirichlet process? 
It can represent any distribution by an infinite mixture of whatever kernel mass 
or density functions you specify. 
It is now straightforward to fit by MCMC.
\end{quote}

What is natural or unnatural is in the eye of the beholder.
I follow Pitman's 1979 eye.
The user of the Dirichlet process has to specify a prior ``concentration" 
parameter -- essentially the density of the number of mixture components -- as 
well as the prior kernel density, and the resulting distribution is sensitive to 
the choice of the concentration parameter.
(A good example of this sensitivity is given in Lunn, Jackson, Best, Thomas and 
Spiegelhalter 2013 pp. 293-296, of the number of mixture components in the 
well-known galaxy data.
Two different settings of the concentration parameter lead to posterior 
distributions with monotonically decreasing probabilities of up to 6 or up to 11 
components.)
 
The Bayesian bootstrap approach depends only on the non-informative Dirichlet
prior, which avoids having to specify anything about the population proportions 
of unobserved values.
The kernel density of the Dirichlet process implies such a specification.

\section{Extensions}

The posterior weighting can be extended straightforwardly to more complex models.
Aitkin (2008 and 2010 \S 4.8) described a two-level model with posterior 
weighting independently within each upper-level unit.
Extensions to more than two levels will require weighting at each level above 
the lowest, but no new features occur.
Multilevel models for large-scale national and international surveys of 
educational attainment, for example the US National Assesment of Educational 
Progress -- the NAEP -- discussed in Aitkin and Aitkin (2011), can be 
generalised in this way.
These and other applications will be described elsewhere. 

\section{Acknowledgements}

In developing this work I am very much indebted to the senior statisticians who 
visited Lancaster during my period there, in particular George Barnard, Jim 
Berger, Darrell Bock, Steve Fienberg, Nan Laird, Richard Royall, Don Rubin and 
David Sprott.

On foundational issues I have benefitted at various times from discussions with 
George Barnard, Peter Green, Bruce Lindsay, Jim Lindsey, Charles Liu, Sylvia 
Richardson, Adrian Smith, David Sprott, Peter Walley and Alan Welsh. 

I am indebted to Jon Cohen, Gary Phillips, Andy Kolstad and many others at the 
American Institutes for Research and the US National Center for Education 
Statistics, and to Jon Rao, Rod Little and Phil Kott, for detailed discussions 
of the survey sampling approach and the design-based/model-based controversy.
I am particularly grateful to Brian Francis for comments on an earlier version, 
which have greatly improved the paper.

This work and its predecessors has been funded over many years by the UK Social 
Science Research Council, the Australian Research Council, and the Institute of 
Education Sciences and the National Center for Education Statistics of the US 
Department of Education.

\section{References}

\begin{description}

\item Aitkin, M. (1978) The analysis of unbalanced cross-classifications (with 
Discussion). {\em Journal of the Royal Statistical Society A}, 141, 195-223.

\item Aitkin, M. (2008) Applications of the Bayesian bootstrap in finite 
population inference. {\em Journal of Official Statistics} 24, 21-51.

\item Aitkin, M., Francis, B., Hinde, J. and Darnell, R. (2009) {\em Statistical 
Modelling in R}. Clarendon Press, Oxford.

\item Aitkin, M. (2010) {\em Statistical Inference: an Integrated 
Bayesian/Likelihood Approach}. Boca Raton, CRC Press.

\item Aitkin, M. and Aitkin I. (2011) {\em Statistical Modeling of 
the National Assessment of Educational Progress}. Springer, New York.

\item Aitkin, M. and Liu, C.C. (2018) Confidence, credibility and prediction 
(with discussion by Little and Welsh and response). {\em Metron} {\bf 76}, 
305-320.

\item Datta, G., Rao, J. and Torabi, M. (2010) Pseudo-empirical Bayes 
estimation of small area means under a nested error linear regression model with 
functional measurement errors. {\em Journal of Statistical Planning and 
Inference} 140, 2952-2962.

\item Ericson, W.A. (1969) Subjective Bayesian models in sampling finite 
populations (with Discussion). {\em Journal of the Royal Statistical Society B} 
31, 195--233.

\item Finney, D. (1947) The estimation from individual records of the 
relationship between dose and quantal response. {\em Biometrika} 34, 320-334. 

\item Guti\'{e}rrez-Pena, E. and Walker, S. (2005) Statistical decision 
problems and Bayesian nonparametric methods. {\em International Statistical 
Review} 73, 309-330.

\item Haldane, J. (1948) The precision of observed values of small frequencies.
\break {\em Biometrika} 35, 297-300. 

\item Hartley, H.O. and Rao, J.N.S. (1968) A new estimation theory for sample 
surveys. {\em Biometrika} 55, 547--557.

\item Hoadley, B. (1969) The compound multinomial distribution and Bayesian 
analysis of categorical data from finite populations. {\em Journal of the 
American Statistical Association} 64, 216--229.

\item Huang, A. (2014) Joint estimation of the mean and error distribution in 
generalized linear models. {\em Journal of the American Statistical Association} 
109, 186-196.

\item Huang, A. (2017) Mean-parametrized Conway-Maxwell-Poisson regression 
models for dispersed counts. {\em Statistical Modelling: an International 
Journal}, 17, 359-380.

\item Lindsey, J.K. (1971) {\em Analysis and Comparison of Some Statistical 
Models}. University of London, PhD thesis.

\item Lindsey, J.K. (1974a) Comparison of probability distributions. {\em 
Journal of the Royal Statistical Society B} 36, 38-47.

\item Lindsey, J.K. (1974b) Construction and comparison of statistical models. 
{\em Journal of the Royal Statistical Society B} 36, 418-425.

\item Lindsey, J.K. (1997) {\em Applying Generalized Linear Models}. New York, 
Springer.

\item Lindsey, J.K and Mersch, G. (1992) Fitting and comparing probability 
distributions with log linear models. {\em Computational Statistics and Data 
Analysis} 13, 373-384.

\item Lunn, D., Jackson, C., Best, N., Thomas, A. and Spiegelhalter, D. (2013)
{\em The BUGS Book: a Practical Introduction to Bayesian Analysis}. CRC Press, 
Boca Raton.

\item Nelder, J. and Wedderburn, R. (1972) Generalized linear models. 
{\em Journal of the Royal Statistical Society A}, 135, 370-384.

\item Owen, A. (1988) Empirical likelihood ratio confidence intervals for a 
single functional. {\em Biometrika} 75, 237-249.

\item Owen, A. (2001) {\em Empirical Likelihood}. Boca Raton: Chapman and 
Hall/CRC.

\item Pitman, E.J.G. (1979) {\em Some Basic Theory for Statistical Inference}.
London: Chapman and Hall. 

\item Rao, J. and Wu, C. (2010a) Pseudo-empirical likelihood inference for 
multiple frame surveys. {\em Journal of the American Statistical Association} 
105, 1494-1503.

\item Rao, J. and Wu, C. (2010b) Bayesian pseudo-empirical likelihood intervals 
for complex surveys. {\em Journal of the Royal Statistical Society B} 72, 
533-544.

\item Rubin, D. (1981) The Bayesian bootstrap. {\em Annals of Statistics} 9, 
130-134.

\item S\"{a}rndal, C.-E., Swensson, B. and Wretman, J. (1992) {\em 
Model-assisted Survey Sampling}. New York: Springer.

\item Smith, T. (1976) The foundations of survey sampling: a review (with 
Discussion). {\em Journal of the Royal Statistical Society A}, 139, 183-204.

\item Tanner, M. and Wong, W. (1987) The calculation of posterior distributions 
by data augmentation. {\em Journal of the American Statistical Association} 82,
528-550.

\item Venables, W. and Ripley, B. (2002) {\em Modern Applied Statistics with S}. 
Fourth edition. Springer.

\item Walker, S. and Guti\'{e}rrez-Pena, E. (2007) Bayesian parametric inference 
in a nonparametric framework. {\em Test} 16, 188-197.
   
\item Wedderburn, R. (1974) Quasi-likelihood functions, generalized linear 
models and the Gauss-Newton method. {\em Biometrika} 61, 439-447. 

\item Wu, C. and Rao, J. (2010) Bootstrap procedures for the pseudo empirical 
likelihood method in sample surveys. {\em Statistics and Probability Letters} 80, 
1472-1478.

\item Yi G.Y., Rao, J.N.S. and Li, H. (2016)  A weighted composite likelihood 
approach for analysis of survey data under two-level models. {\em Statistica 
Sinica} 26, 569-587.

\item Zhang, N. and Huang, A. (2018) Profile likelihood ratio tests for 
parameter inferences in generalised single-index models. {\em Journal of 
Nonparametric Statistics}. doi:10.1080/10485252.2018.1506121

\end{description}
\end{document}